\documentclass[a4paper,twoside,psfig]{article}
\newcommand{\affil}[1]{$^{\rm #1}$}
\usepackage{graphicx}

\date{} 

\newcommand{\HI}{H\,{\sc i}}
\newcommand{\kms}{~km\,s$^{-1}$}
\newcommand{\kkms}{km\,s$^{-1}$}
\newcommand{\degr}{^{\circ}}

\newcommand{\vopt}{$v_{\rm opt}$}
\newcommand{\vsys}{$v_{\rm sys}$}
\newcommand{\tsys}{$T_{\rm sys}$}

\newcommand{\FHI}{$F_{\rm HI}$}
\newcommand{\MHI}{$M_{\rm HI}$}
\newcommand{\Msun}{~M$_{\odot}$}

\title{\large\bf\flushleft 
  Overview on spectral line source finding and visualisation}

\author{\parbox{\textwidth}{\flushleft
\vspace{-0.5cm}
{\it B\"arbel S. Koribalski\affil{A,B}} \\
\vspace{0.4cm}
{\small \affil{A}\,CSIRO Astronomy \& Space Science, 
    Australia Telescope National Facility, P.O. Box 76,
    Epping, NSW 1710, Australia}\\
{\small \affil{B}\,Email: Baerbel.Koribalski@csiro.au}}}
\begin{document}
\twocolumn[
\begin{changemargin}{.8cm}{.5cm}
\begin{minipage}{.9\textwidth}
\vspace{-1cm}
\maketitle
%
%

\small{\bf Abstract:}

Here I will outline successes and challenges for finding spectral line sources 
in large data cubes that are dominated by noise. This is a 3D challenge as the 
sources we wish to catalog are spread over several spatial pixels and spectral 
channels. While 2D searches can be applied, e.g., channel by channel, optimal
searches take into account the 3-dimensional nature of the sources. In
this overview I will focus on \HI\ 21-cm spectral line source detection in 
extragalactic surveys, in particular HIPASS, the {\em HI Parkes All-Sky Survey}
and WALLABY, the {\em ASKAP HI All-Sky Survey}. I use the original HIPASS data
to highlight the diversity of spectral signatures of galaxies and gaseous 
clouds, both in emission and absorption. Among others, I report the discovery 
of a 680\kms\ wide \HI\ absorption trough in the megamaser galaxy NGC~5793. 
Issues such as source confusion and baseline ripples, typically encountered 
in single-dish \HI\ surveys, are much reduced in interferometric \HI\ surveys.
Several large \HI\ emission and absorption surveys are planned for the 
Australian Square Kilometre Array Pathfinder (ASKAP): here we focus on WALLABY,
the 21-cm survey of the sky ($\delta < +30\degr$; $z < 0.26$) which will take
about one year of observing time with ASKAP. Novel phased array feeds (``radio 
cameras") will provide 30 square degrees instantaneous field-of-view. WALLABY 
is expected to detect more than 500\,000 galaxies, unveil their large-scale 
structures and cosmological parameters, detect their extended, low-surface 
brightness disks as well as gas streams and filaments between galaxies. It is 
a precursor for future \HI\ surveys with SKA Phase\,I and II, exploring galaxy 
formation and evolution. The compilation of highly reliable and complete source 
catalogs will require sophisticated source-finding algorithms as well as 
accurate source parametrisation. 



\medskip
\medskip
\end{minipage}
\end{changemargin}
]
\small

\section{Introduction} 

In recent years many remarkable galaxy surveys at optical and infrared 
wavelengths have become available. The Sloan Digital Sky Survey (SDSS; York
et al. 2000) databases currently hold many millions of galaxies and nearly 
one million optical spectra (Eisenstein et al. 2011). McConnachie et al. 
(2009) extracted 29 million galaxies from SDSS DR6 to study Hickson Compact 
Groups and identified nearly 400 000 galaxy groups. Over 1.5 million galaxies 
are listed in the 2MASS Extended Source Catalog (XSC; Jarrett et al. 2000), 
and a WISE Extended Source Catalog has recently been released (Jarrett et al. 
2012). Multi-colour optical sky surveys such as PanSTARRS (Stubbs et al. 2010) 
and SkyMapper (Keller et al. 2007) are under way. 

The NASA Extragalactic Database (NED) currently holds over 100 million 
objects classified as galaxies (Marion Schmitz, priv. comm.), and the Lyon 
Extragalactic Database (LEDA) contains at least 1.5 million bona-fide 
galaxies. In comparison, the total number of \HI\ 21-cm detected galaxies is 
small: several tens of thousands (Meyer et al. 2004, Springob et al. 2005, 
Wong et al. 2006, Haynes et al. 2011). The intrinsic faintness of the electron 
spin-flip transition of neutral atomic hydrogen (rest frequency 1.42~GHz) 
makes it difficult to detect \HI\ emission from individual galaxies at large 
distances. To study the \HI\ content of galaxies and diffuse \HI\ filaments 
between galaxies, we need radio synthesis telescopes with large collecting 
areas, low-noise receivers and large fields of view. Such requirements 
provide a range of engineering and computing challenges (Chippendale et al. 
2010, Cornwell 2007, Schinckel et al. 2012). 

While we have come a long way since the discovery of the 21-cm spectral line 
by Ewen \& Purcell in 1951, detecting a Milky Way-like galaxy at redshift $z 
= 1$ in \HI\ emission will require the Square Kilometre Array (Obreschkow et 
al. 2011). Several large radio surveys have been proposed for SKA pathfinder 
and precursor telescopes and are currently undergoing intense design studies. 
The latter include evaluating and improving source-finding algorithms which 
are needed to extract scientifically useful (i.e., highly reliable and 
complete) source catalogs from the large survey volumes.

This paper is organised as follows: in Section~2 I will briefly introduce 
the SKA Pathfinder \HI\ All-Sky Surveys (WALLABY and WNSHS), followed by an
overview on source finding issues and algorithms in Section~3. In Section~4 
I use the {\em HI Parkes All-Sky Survey} (HIPASS; Barnes et al. 2001, 
Koribalski et al. 2004) to illustrate the diversity of galaxy \HI\ profiles
and present new detections (see Figs.~1--3). An overview of 3D visualisation 
tools, to explore source catalogs and data cubes, is given in Section~5. 

\begin{figure*}[ht]
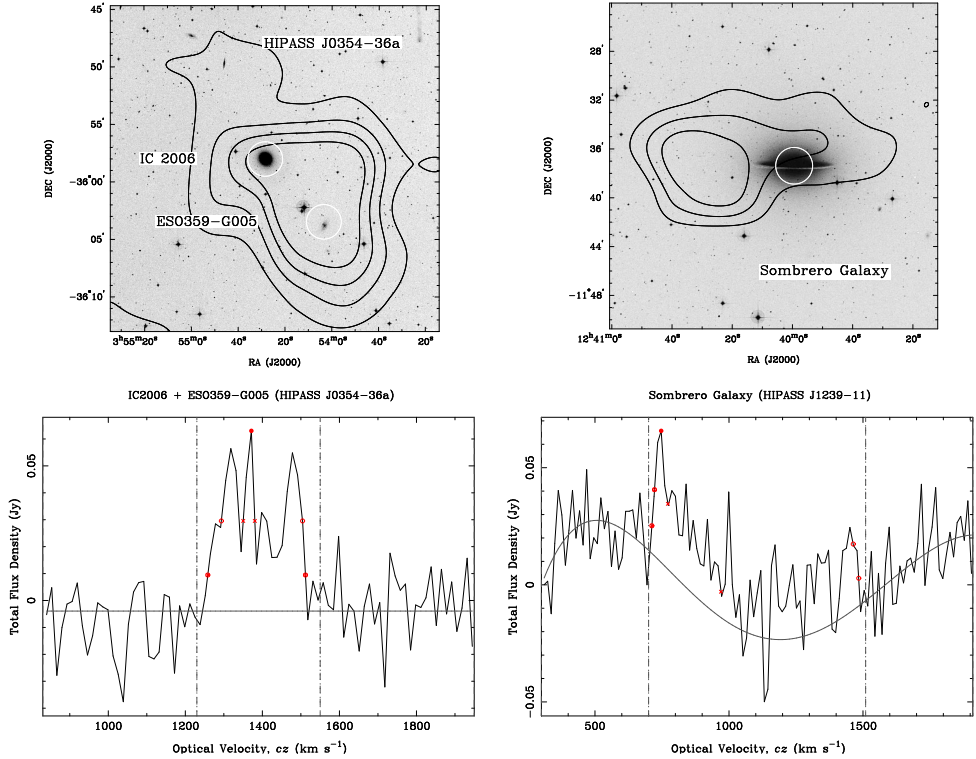
 
\centering
\begin{tabular}{cc}
 \includegraphics[scale=0.25,angle=-90]{HIPASSJ0354-36a.mom0.regrid.ps}  &
 \includegraphics[scale=0.25,angle=-90]{HIPASSJ1239-11.mom0.regrid.ps}  \\ 
 \includegraphics[scale=0.25,angle=-90]{HIPASSJ0354-36a.spec.ps} &
 \includegraphics[scale=0.25,angle=-90]{HIPASSJ1239-11.spec.ps} \\
\end{tabular}
\caption{\small
   Integrated \HI\ distribution overlaid on DSS $B$-band images (top) 
   and \HI\ spectra (bottom) for two galaxy systems: {\bf HIPASS J0354--36a} 
   (left) and {\bf HIPASS J1239--11} (right). The data shown here are from the 
   {\em HI Parkes All-Sky Survey} (HIPASS); the gridded beam is 15.5 arcmin.
   HIPASS J0354--36a encompasses 
   the S0 galaxy IC\,2006 and the dwarf galaxy ESO359-G005; the \HI\ contour
   levels are 1, 2, 3 and 4 Jy\,beam$^{-1}$\kms. HIPASS J1239--11 is better 
   known as the Sombrero Galaxy (M\,104); the \HI\ contour levels are 2, 3 and 
   4 Jy\,beam$^{-1}$\kms. Optical galaxy positions are indicated with circles. 
   Red markers in the HIPASS spectra indicate fitted \HI\ properties such as 
   the peak flux and the 50\% and 20\% velocity width. The fitted baseline is 
   shown in grey (0th order for HIPASS J0354--36a and 5th order for HIPASS 
   J1239--11).}
\label{fig:figure1}
\end{figure*}

\begin{figure*}[ht]
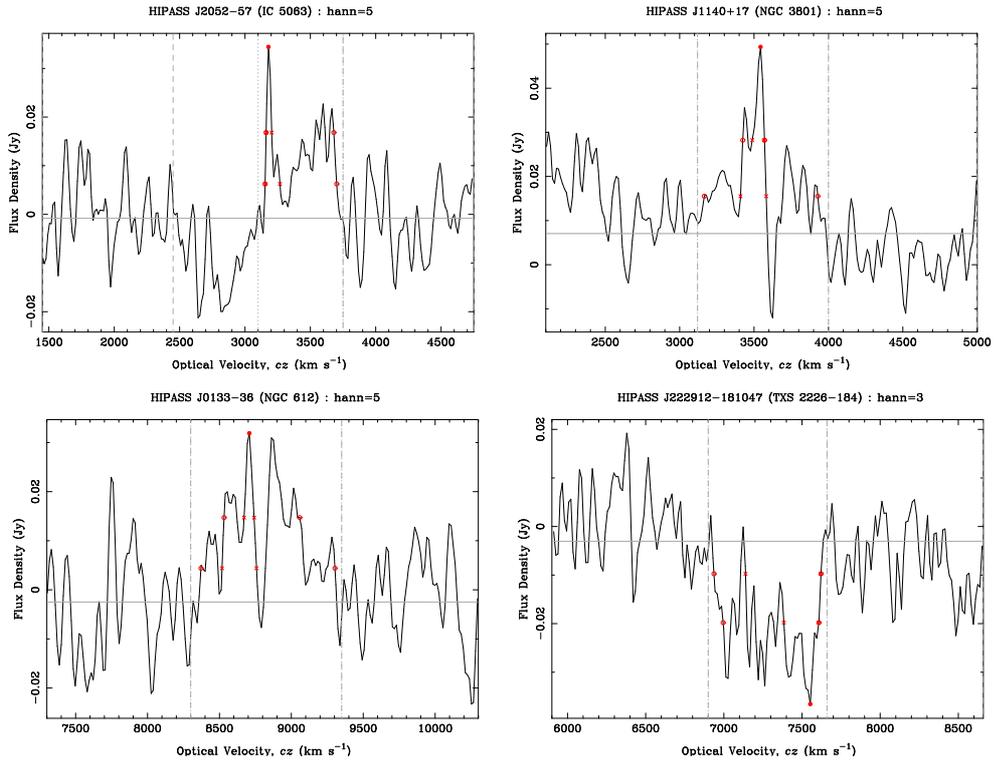
 
\centering
\begin{tabular}{cc}
 \includegraphics[scale=0.25,angle=-90]{ic5063.hann5.ps}   &
 \includegraphics[scale=0.25,angle=-90]{ngc3801.hann5.ps}  \\
\vspace{0.5cm}
 \includegraphics[scale=0.25,angle=-90]{ngc612.hann5.ps}   &
 \includegraphics[scale=0.25,angle=-90]{txs2226-184.hann3.ps}   \\
\end{tabular}
\caption{\small
   HIPASS spectra of the galaxies IC\,5063 (top left), NGC~3801 (top
   right), NGC~612 (bottom left) and TXS 2226--184 (bottom right). 
   Hanning smoothing was used to improve the signal to noise of the detected
   \HI\ emission and absorption features (5-point Hanning = 52\kms; 3-point
   Hanning = 26\kms). Fitted \HI\ properties are indicated by red markers in 
   the HIPASS spectra; the fitted 0th order baseline is shown in grey.}
\label{fig:hiabs}
\end{figure*}

\begin{figure*}[ht]
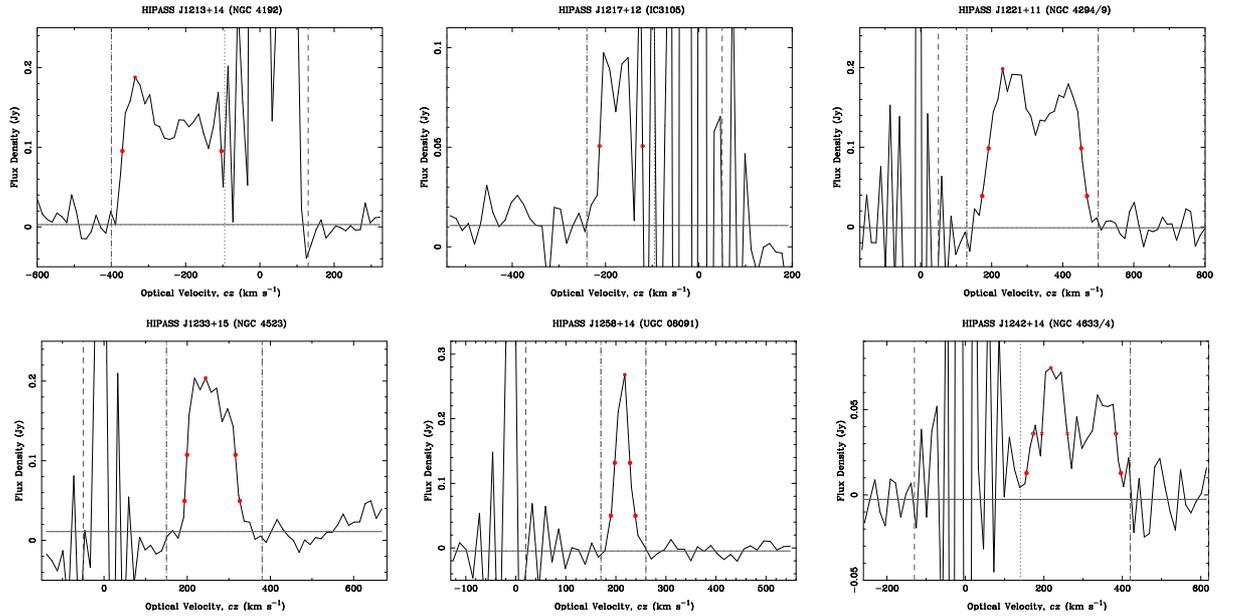
 
\centering
\begin{tabular}{ccc}
 \includegraphics[scale=0.2,angle=-90]{HIPASSJ1213+14.spec.ps}   &
 \includegraphics[scale=0.2,angle=-90]{HIPASSJ1217+12.spec.ps}   &
 \includegraphics[scale=0.2,angle=-90]{HIPASSJ1221+11.spec.ps}   \\
\vspace{0.5cm}
 \includegraphics[scale=0.2,angle=-90]{HIPASSJ1233+15.spec.ps}   &
 \includegraphics[scale=0.2,angle=-90]{HIPASSJ1258+14.spec.ps}   &
 \includegraphics[scale=0.2,angle=-90]{HIPASSJ1242+14.spec.ps}   \\
\end{tabular}
\caption{\small
   HIPASS spectra of the Virgo cluster members NGC~4192, IC\,3105, 
   NGC~4594/9 (top row, from left to right), NGC~4523, UGC~08091, and 
   NGC~4633/4 (bottom row, from left to right). No Hanning smoothing 
   was used here. Bright Galactic \HI\ emission is seen as artifacts 
   in all displayed spectra. Fitted \HI\ properties are indicated by 
   red markers in the HIPASS spectra; the fitted 0th order baseline 
   is shown in grey.}
\label{fig:hiabs}
\end{figure*}

\begin{table*} 
\begin{tabular}{lccc}
\hline
\hline
             & \\
             & {\bf HIPASS}        & {\bf WALLABY}       & {\bf WNSHS}   \\
             &     (1)             &     (2)             &     (3)       \\
\hline
\hline
             & \\
telescope    & 64-m {\bf Parkes} dish & {\bf ASKAP}      & {\bf WSRT}        \\
             &         & $36 \times 12$-m dishes & $12 \times 25$-m dishes   \\
             &         & {\em (under construction)}      &  (maxi-short)     \\
baselines    & ---                 & 20 m to 2 (6) km & 36 m to $\sim$2.5 km \\
\hline
             & \\
receiver     & 21-cm multibeam     & phased array feed   & phased array feed \\
             &                     & (Chequerboard)      & (Vivaldi)     \\
\tsys\ spec. & 20 K                & 50 K                & 50 K          \\
field-of-view& $\sim$1 deg$^2$ (13 beams) & 30 deg$^2$   & 8 deg$^2$     \\
obs. mode    & scanning            & dithering/mosaicing & mosaicing     \\
angular resolution  & 15.5 arcmin  & 30 (10) arcsec      & 30 (15) arcsec  \\
pixel size   & 4 arcmin            & 7.5 (2.5) arcsec    & 10 (4.5) arcsec \\
\hline
             & \\
sky coverage & $\delta < +25\degr$ & $\delta < +30\degr$ & $\delta > +27\degr$\\
             & 29 343 deg$^2$      & 30 940 deg$^2$      & 11 262 deg$^2$ \\
cubes/fields & 538 ($8\degr \times 8\degr$) & 1300 & 1410 \\
\hline
                    & \\
bandwidth           & 64 MHz       & \multicolumn{2}{c}{300 MHz}     \\
no. of channels     & 1024         & \multicolumn{2}{c}{16384}       \\
channel width       & 13.2\kms     & \multicolumn{2}{c}{3.9\kms}     \\
velocity resolution & 18.0\kms     & \multicolumn{2}{c}{3.9\kms}     \\
\hline
                    & \\
integration time    & 450~s        & 8~h          & 4 (12) h     \\
~~~per pointing     &              &              &              \\
rms per channel     & $\sim$13 mJy\,beam$^{-1}$ 
                    & 1.6 mJy\,beam$^{-1}$      
                    & 1.5 (1.2) mJy\,beam$^{-1}$ \\
frequency coverage  & 1362.5 -- 1426.5 MHz 
                      & \multicolumn{2}{c}{1130 -- 1430 MHz} \\
velocity range ($cz$) & --1280 to +12 700\kms 
                      & \multicolumn{2}{c}{--2000 to +77 000\kms} \\
               & $z < 0.04$        & \multicolumn{2}{c}{$z < 0.26$} \\
galaxies     & $\sim$6500          & $\sim$500 000  & $\sim$100 000 \\
duration     & $\sim$1997 -- 2002  & \multicolumn{2}{c}{from 2014?} \\
\hline
\hline
\end{tabular}
\caption{Comparison of \HI\ survey parameters for HIPASS, WALLABY and WNSHS. 
     The parameters for WALLABY and WNSHS are approximate and may change in 
     future. {\bf Notes:} (1) Barnes et al. (2001) and references therein;
     (2) Koribalski \& Staveley-Smith (2009), see 
         {\em www.atnf.csiro.au/research/WALLABY}; 
     (3) J\'ozsa et al. (2010), see {\em www.astron.nl/$\sim$jozsa/wnshs}. 
         Details are given in Section~2.} 
\end{table*}

\section{SKA Pathfinder HI Surveys} 

ASKAP, the {\em Australian SKA Pathfinder} (DeBoer et al. 2009), consists of
$36 \times 12$-m antennas and is located in the Murchinson Shire of Western
Australia. As of June 2012 the construction of all ASKAP antennas is completed.
Of the 36 antennas, 30 are located within a circle of $\sim$2~km diameter,
while six antennas are at larger distances providing baselines up to 6~km. 
Each ASKAP dish will be equipped with novel Chequerboard phased array feeds 
(PAFs). We expect the first six Mk1 PAFs (the co-called BETA array) to be 
ready for engineering testing by the end of 2012, with science commissioning 
to be integrated when feasible. The instantaneous field-of-view of the ASKAP 
PAFs is 5.5 deg $\times$ 5.5 deg, ie. 30 square degrees (Chippendale et al. 
2010), making ASKAP a 21-cm survey machine. The WSRT APERTIF upgrade employs 
Vivaldi PAFs, delivering a field-of-view of eight square degrees (Verheijen 
et al. 2008). The system temperature specification of the ASKAP PAFs is 50~K 
over the full band, with future PAF generations promising much higher 
performance.  \\

{\bf WALLABY}, the {\em Widefield ASKAP L-band Legacy All-sky Blind surveY}
(led by B\"arbel Koribalski \& Lister Staveley-Smith; see Koribalski et al. 
2009), will cover 75\% of the sky ($-90\degr < \delta < +30\degr$) over a 
frequency range from 1.13 to 1.43~GHz (corresponding to $-2000 < cz < 
77,000$\kms) at resolutions of $30''$ and 4\kms. For further details see 
Table~1. WALLABY will be carried out using the inner 30 antennas of ASKAP,
which provide excellent $uv$-coverage and baselines up to 2~km. 
High-resolution ($10''$) ASKAP \HI\ observations using the full 36-antenna 
array will require further computing upgrades.

{\bf WNSHS}, the {\em Westerbork Northern Sky HI Survey} (led by Guyla J\'ozsa;
see J\'oza et al. 2010), will cover a large fraction of the northern sky 
($\delta > +27\degr$) with APERTIF (Verheijen et al. 2008) over the same 
frequency range as WALLABY with ASKAP. Both \HI\ surveys combined will 
achieve a true all-sky survey with unprecedented resolution and depth. The 
science goals of both surveys are well developed and complement, as well 
as enhance, each other. 

WALLABY and WNSHS are made possible by the development of phased array feeds, 
delivering a much larger field-of-view than single feed horns or multi-beam 
systems. A comparison of \HI\ survey parameters for HIPASS, WALLABY, and 
WNSHS is given in Table~1.

WALLABY will take approximately one year (ie 8 hours per pointing) and deliver 
an rms noise of 1.6 mJy\,beam$^{-1}$ per 4\kms\ channel. We estimate that more 
than 500\,000 gas-rich galaxies can be individually detected within the WALLABY
volume (Johnston et al. 2008, Duffy et al. 2012, Koribalski et al. 2012). The 
WALLABY team will examine the \HI\ properties and large-scale distributions of 
galaxies out to a redshift of $z = 0.26$ (equivalent to a look-back time of 
$\sim 3$ Gyr) in order to study: (1) galaxy formation and the missing satellite
problem in the Local Group, (2) evolution and star-formation in galaxies, (3) 
mergers and interactions in galaxies, (4) the \HI\ mass function and its 
variation with galaxy density, (5) physical processes governing the 
distribution and evolution of cool gas at low redshift, (6) cosmological 
parameters relating to gas-rich galaxies, and (7) the nature of the cosmic web. 
WALLABY will detect dwarf galaxies (\MHI\ = 10$^8$\Msun) out to a distance of 
$\sim$60 Mpc, massive galaxies ($M_{\rm HI}^* = 6 \times 10^9$\Msun) to 
$\sim$500 Mpc, and super-massive galaxies like Malin\,1 (\MHI\ = $5 \times 
10^{10}$\Msun) to the survey `edge' of 1~Gpc. The mean sample redshift is 
expected to be $z$ = 0.05 (200~Mpc).

\section{Source finding considerations} 

Reliability and completeness are of high importance when compiling source 
catalogs as knowing both is essential for the statistical analysis and 
interpretation of source properties. The process of {\em finding sources} 
can be considered one of many important steps (e.g., data pre-processing, 
source finding \& characterisation, cataloging / post-processing) in the 
production of astronomical source catalogs.

The largest data volumes will come from wide-field spectral line surveys 
such as WALLABY and WNSHS, while radio continuum and polarisation surveys are 
typically an order of magnitude smaller. Apart from GASKAP, the {\em Galactic 
ASKAP Survey}\footnote{Within their proposed 7.3~MHz band, centred on the 
   Galactic \HI\ line, GASKAP will also be able to detect nearby, gas-rich 
   galaxies at high velocity resolution (0.25\kms). This is of great benefit 
   to the kinematical study of low-mass dwarf galaxies which rotate slowly 
   and often display high turbulence and non-circular motions.}
(Dickey et al. 2012), ASKAP 21-cm surveys will use the full 300~MHz bandwidth. 
To achieve $\sim$4\kms\ velocity resolution, the extragalactic \HI\ surveys, 
WALLABY, WNSHS and also the {\em Deep Investigations of Neutral Gas Origins} 
(DINGO) survey, require the 300~MHz bandwidth to be divided into 16384 
channels. 

In contrast, radio continuum surveys such as EMU, the {\em Evolutionary Map 
of the Universe} survey (Norris et al. 2011), use the same band divided into 
300 channels, each 1~MHz wide. EMU and WALLABY are likely to jointly 
survey the sky at 21-cm. While the WALLABY team expects to individually detect 
more than half a million gas-rich galaxies within the survey volume (3.26 
Gpc$^3$; see Section~2), the EMU team expects to detect $\sim$70 million 
sources. 

Huynh et al. (2012) explore 2D source finding algorithms such as {\sc SFind} 
(Hopkins et al. 2002), {\sc S-Extractor} (Bertin \& Arnouts 1996), and {\sc 
Duchamp} (Whiting 2008; Whiting 2012), optimised for compact continuum sources. 
The high continuum source density means that confusion, dynamic range, and 
source identification (which is essential to gather optical redshifts) are key 
issues. Hancock et al. (2012) look into compact continuum source-finding and
compare {\sc S-Extractor}, {\sc MIRIAD} {\sc imsad} (Sault et al. 1995), {\sc 
SFind} and {\sc Aegean}. The {\em Circle Hough Transform} is explored by 
Hollitt \& Johnston-Hollitt (2012) for the detection of extended and diffuse 
objects such as supernova remnants, radio galaxies and relics. Molinari et 
al. (2011) look into source extraction and photometry for continuum surveys 
at mid- and far-infrared as well as sub-millimetre wavelengths. They present 
a new method, {\sc CUTEX} for {\em CUrvature Thresholding EXtractor}, to 
detect sources in the presence of intense and highly variable fore/background 
emission, in particular in the Galactic Plane (see also Marsh \& Jarrett 2012).

Neither source confusion nor dynamic range are concerns for WALLABY. But, as 
the 21-cm line of atomic neutral hydrogen is intrinsically very faint, finding 
and characterising \HI\ clouds, filaments, and galaxies of various sizes in 
the 3D data cubes, is difficult. The 3D shape of \HI\ sources --- in WALLABY 
galaxies are always extended over numerous velocity channels --- typically 
provides a `contiguous block of voxels' which is distinguishable from white 
noise. The \HI\ emission line traces the warm neutral hydrogen gas in galaxies
whose observed velocity dispersion is typically $10 \pm 2$\kms\ (Tamburro et 
al. 2009). This means that an \HI\ emission signal from a galaxy (even a 
face-on galaxy; see Petric \& Rupen 2007) will always extend over at least 
two 4\kms\ wide channels and generally over more than three channels. The 
ASKAP synthesized beam ($\sim 10''/30''$ for the 6-/2-km arrays, respectively) 
is then sampled by at least three pixels in each spatial direction. \\

{\sc Duchamp} (Whiting 2008; Whiting 2012) is one of a number of programs
for finding and characterising astronomical sources in images and data cubes.
For ASKAP 21-cm data a specific version of {\sc Duchamp}, called {\sc Selavy}, 
is being developed in extensive consultation with the ASKAP Survey Science 
Projects (see Whiting \& Humphreys 2012). Basic testing of {\sc Duchamp} using 
(1) a set of spatially unresolved sources with narrow Gaussian spectra and 
(2) a set of spatially resolved sources with double-horn spectra was carried 
out by Westmeier, Popping \& Serra (2012). To improve both the reliability and 
completeness of source finding over a broad range of parameters, a range of
algorithms, based on wavelets, Kuiper and Kolmogorov-Smirnov tests, Bayesian 
statistics, etc., are being explored. 

Among the new 3D algorithms are the {\em Characterised Noise HI} ({\sc CNHI}) 
source-finder (Jurek 2012), the {\em 2D--1D wavelet reconstruction} (Fl\"oer 
\& Winkel 2012), and the {\em Smooth \& Clip} (S+C) technique (Serra et al. 
2012; Popping et al. 2012). A combination of these 
techniques is currently being investigated. Serra, Jurek \& Fl\"oer (2012) 
look into the reliability of source finding algorithms using a statistical 
analysis of the noise characteristics in \HI\ data cubes. Pre-processing 
techniques such as iterative median smoothing and wavelet de-noising are 
discussed by Jurek \& Brown (2012). A comprehensive comparison of the above 
mentioned source-finding algorithms has been conducted by Popping et al. 
(2012) using a range of thresholds and smoothing options. 

Overall, the development of efficient and reliable 3D source finding tools 
will provide a major advance to the analysis of any large spectral line data 
set. It is very important to be able to search data cubes without adopting
any prior knowledge of source properties (e.g., galaxy positions from optical
or infrared catalogs). Such an open approach ensures a large variety of 
sources (in the \HI\ context, these are different types of galaxies and \HI\ 
filaments / clouds) is catalogued and is essential to avoid bias towards a 
narrow range of sources.  Targeted searches, based on prior knowledge of the 
respective sources, would complement the open approach and address specific 
science goals.

Some surveys require highly specialised algorithms, for example searching for 
maser emission in HOPS, the {\em H$_2$O Galactic Plane Survey} (Walsh et al. 
2012), detection thresholds and bias correction for polarised continuum sources
(George, Stil \& Keller 2012), as well as finding and characterising sources 
in 4-colour data from WISE, the {\em Wide-field Infrared Survey Explorer} 
(Marsh \& Jarrett 2012). Radio transients require yet other considerations;
Keith et al. (2010) describe sophisticated pulsar searches in single-dish 
surveys, while Bannister \& Cornwell (2011) introduce two new algorithms for 
the detection of transients in interferometric surveys.

Targeted searches may focus on particular spatial and/or spectral source shapes 
(e.g., very wide double-horn profiles, extended filaments or equi-distant 
recombination lines) using highly optimised search algorithms with build-in
assumptions (so-called priors). For example, Allison et al. (2012a,b) use 
Bayesian statistics to search for \HI\ absorption lines in spectra towards 
bright radio continuum sources. This will be of great importance for SKA 
pathfinder surveys like FLASH, the {\em First Large Absorption Survey in HI} 
led by E. Sadler. Another example is provided by Hurley-Walker et al. (2012) 
who use Bayesian analysis with specific priors to search for the SZ effect 
from galaxy clusters.  \\

The developments described above are targeted at optimising source detection 
for WALLABY and WNSHS. They will also be useful for deep interferometric \HI\ 
surveys, such as DINGO (led by M. Meyer), and LADUMA (led by S. Blyth \& B. 
Holwerda), single-dish \HI\ surveys, such as HIPASS (Barnes et al. 2001; 
Koribalski et al. 2004), HIJASS (Lang et al. 2003; Wolfinger et al. 2012), 
and EBHIS (Winkel et al. 2010), and many other large-volume \HI\ surveys. 

\section{HIPASS source-finding} 

The development of a powerful 13-beam receiver system ($T_{\rm sys} \approx 
20$~K) plus versatile correlator on the 64-m Parkes telescope instigated an
era of large-scale 21-cm surveys of our Galaxy and the Local Universe. The 
{\em HI Parkes All-Sky Survey} (HIPASS) is the largest and most prominent 
of the Parkes \HI\ surveys. It covers the whole sky to a declination limit 
of $\delta = 25\degr$ over a velocity range from --1280 to 12700\kms. The 
Parkes gridded beam is $\sim$15.5 arcmin (to sample the beam adequately a 
pixel size of 4 arcmin is used), the velocity resolution is 18\kms, and the 
rms noise is $\sim$13 mJy\,beam$^{-1}$ per channel (for a typical integration 
time of 8 minutes). See Barnes et al. (2001) for a detailed description of 
the HIPASS observations, calibration and imaging techniques. 

Current galaxy catalogs include the {\em HIPASS Bright Galaxy Catalog} (HIPASS 
BGC; Koribalski et al. 2004; Ryan-Weber et al. 2002; Zwaan et al. 2003), the 
southern HIPASS catalog (HICAT; Meyer et al. 2004) and the northern HIPASS 
catalog (NHICAT; Wong et al. 2006). Together these catalogs, which are highly 
reliable (Zwaan et al. 2004), contain more than 5000 \HI-rich galaxies and a 
few \HI\ clouds (e.g., HIPASS J0731--69; Ryder et al. 2001). In addition,
Parkes \HI\ multibeam surveys of the Zone of Avoidance (ZOA) have catalogued 
more than 1000 galaxies (Staveley-Smith et al. 1998; Henning et al. 2000; 
Juraszek et al. 2000; Donley et al. 2005). Compact and extended populations of 
Galactic high-velocity clouds (HVCs) were catalogued by Putman et al. (2002). 

Because our aim was to produce highly reliable HIPASS catalogs, many faint 
\HI\ sources have not (yet) been catalogued (see Figs.~1 \& 2). Furthermore, 
gas-rich galaxies with velocities less than $\sim$300\kms\ lie outside the 
parameter space considered for (N)HICAT (see Fig.~3). This limit was chosen 
to avoid confusion with Galactic HVCs. 

An advanced HIPASS data reduction is under way (Calabretta et al. 2012, in 
prep.), aiming to reduce on- and off-source spectral ripple as well as improve 
the bandpass calibration, survey sensitivity and source parametrisation. Our 
goal is to obtain much deeper HIPASS catalogs by employing the sophisticated
source-finding algorithms discussed in this {\em PASA Special Issue}. 
A comparison of the original HIPASS data with the new version for several 
areas is under way. 

\subsection{New HIPASS sources} 

In this section, I use HIPASS data to highlight the diversity of spectral 
signatures of galaxies and gaseous clouds, both in emission and absorption.
I present some previously uncatalogued HIPASS detections of galaxies. Their 
low signal-to-noise and/or low velocity prevented inclusion in the published
catalogs which were compiled following a {\em blind} search and verification
based on the HIPASS data alone (see Meyer et al. 2004; Wong et al. 2006).
The new HIPASS detections of galaxy groups, pairs and individual spirals are 
shown in Figs.~1--3. HIPASS names are assigned as previously, using their 
fitted \HI\ position. Detailed source descriptions are given in the Appendix, 
and their HIPASS properties are summarised in Table~2. \\

Galaxies with notable \HI\ absorption features detected in HIPASS are briefly
discussed by Koribalski et al. (2004; Section~3.6). Prominent examples are 
NGC~253 (HIPASS J0047--25), NGC~3256 (HIPASS J1028--43), NGC~4945 (HIPASS 
J1305--49), Circinus (HIPASS J1413--65) and NGC~5128 (HIPASS J1324--42). Some 
galaxies with bright radio nuclei, such as NGC~5793 (HIPASS J1459--16A), are 
easily detected in \HI\ absorption but not seen in \HI\ emission, highlighting 
the fact that such galaxies would be missing from any \HI\ peak-flux limited 
catalogs despite their substantial \HI\ content. Because of the large Parkes 
beam (15.5 arcmin) HIPASS spectra of individual nearby galaxies with bright 
radio continuum emission may show a combination of \HI\ emission from the 
galaxy disk and \HI\ absorption against the star-forming nuclear region
(examples are shown in Fig.~2 and discussed in the Appendix).  \\

Here I report the discovery of a 680\kms\ wide \HI\ absorption trough in the 
megamaser galaxy NGC~5793. This feature is seen in addition to the well-known
narrow \HI\ absorption line reported by Pihlstr\"om et al. (2000). Figs.~4 
\& 5 show the respective HIPASS spectra; further details are given in the 
Appendix. \\

Galaxies in the Virgo cluster have systemic velocities typically ranging from 
--700 to +2700\kms. Notably, at velocities around zero, single-dish \HI\ 
observations are confused by Galactic \HI\ emission and HVCs, and the galaxy
\HI\ properties can be hard to measure accurately. The HIPASS spectra of nine 
Virgo galaxies are shown in Fig.~3 and their properties are summarised in
Table~2. Independent distance estimates to Virgo galaxies range from about 16 
to 24 Mpc. 

\begin{figure}[ht] 
 \includegraphics[scale=0.3,angle=-90]{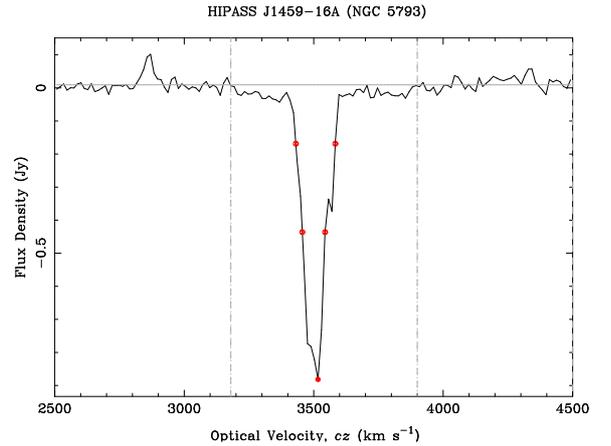} 
\caption{\small
  HIPASS spectrum towards the Sy\,2 galaxy NGC~5793 (\vopt\ = 3491\kms), 
  also known as PKS~1456--164 ($\sim$1 Jy at 1.4~GHz) and HIPASS J1459--16A
  (Koribalski et al. 2004). We detect the well-known deep \HI\ absorption 
  feature ($\sim$3420 to 3590\kms) and a very wide \HI\ absorption trough
  ($\sim$3200 to 3880\kms). For a detailed view of the latter see Fig.~5. 
  --- The \HI\ emission feature at 2861\kms\ (HIPASS J1459--16B) is associated 
  with the dwarf irregular galaxy 6dF J1459410--164235 (\vopt\ = 2857\kms) and 
  not the E0 galaxy NGC~5796 (\vopt\ = 2971\kms).}
\label{fig:hiabs}
\end{figure}

\begin{figure}[ht] 
 \includegraphics[scale=0.3,angle=-90]{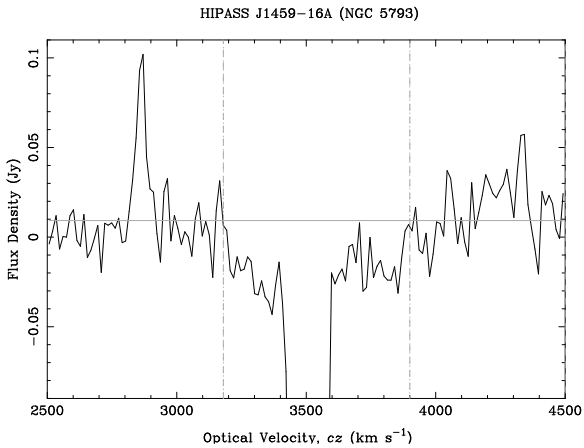} 
\caption{\small
  A closer look at the megamaser galaxy NGC~5793 reveals a very broad 
  \HI\ absorption feature, extending from $\sim$3200 to 3880\kms\ or --290 to 
  +390\kms\ with respect to NGC~5793's systemic velocity. This remarkable 
  680\kms\ wide feature has not previously been detected and requires further 
  investigation.}
\label{fig:ngc5793-zoom}
\end{figure}

\subsection{Statistical techniques} 

To study the properties of astronomical sources below the survey detection 
threshold, statistical approaches such as stacking and intensity mapping 
are used. For example, Pen et al. (2009) use the HIPASS data to co-add \HI\ 
spectra at the positions of 27417 optical galaxies, after shifting their 
systemic velocities to a common restframe. As expected, this results in a 
large \HI\ signal, dominated by the bright \HI\ emission of individually 
detected galaxies. After removing all galaxies listed in HICAT (Meyer et al. 
2004) the co-added \HI\ signal is detected at high significance (Meyer, 
priv. comm.). The deep ASKAP \HI\ survey DINGO will detect most \HI\ sources 
through stacking of spectra from already known galaxies with accurate 
velocities).

Another technique to study \HI\ emission in and between galaxies over large 
volumes is known as {\em intensity mapping}. Instead of aiming to individually 
detect galaxies, which requires high angular resolution and sensitivity, this 
approach can be employed to measure the collective emission of many galaxies 
at low angular resolution (several tens of Mpc). The 21-cm {\em intensity 
mapping} allows the 3D measurement of large scale structures and velocity/flow 
fields to large redshifts. 

\section{Visualisation} 

Visual data exploration and discovery is used across all sciences with a 
large range of tools and algorithms available. Each discipline requires 
suitable software to analyse their data, the complexity and volume of which
is growing steadily. Here I look into astronomical software packages and 
specific algorithms that allow the visualisation of galaxy data (incl. 
numerical simulations) from individual objects to survey catalogs.

3D visualisations of several thousand catalogued HIPASS galaxies (see 
Section~4) were created by Mark Calabretta and are available 
on-line\footnote{\em www.atnf.csiro.au/people/mcalabre/animations}.
The animations depict the distribution of gas-rich galaxies in the nearby
Universe ($z < 0.03$), taking into account their positions, velocities /
distances and \HI\ masses. Large-scale structures such as the Supergalactic 
Plane and the Local Void (see Koribalski et al. 2004) are clearly visible. In
future we should be able to replace each source, currently depicted by a 
sphere, by a 3D multi-wavelength rendering of the respective galaxy. 

Dolag et al. (2008) outline the wide-ranging benefits of data visualisation, 
in particular 3D rendering of complex data sets, and introduce the {\sc 
SPLOTCH} public ray-tracing software (see description below). They emphasize 
the need for tools to visualise scientific data in a {\em comprehensive, 
self-describing, rich and appealing way}. {\sc SPLOTCH} is a powerful
package to do this (see also Jin et al. 2010). \\

Multi-wavelength images and spectral line data cubes of galaxies allow us to
measure their stellar, gas and dark matter properties. Visualisation packages
such as {\sc KARMA} (see below) provide a range of tools to interactively view
2D and 3D data sets as well as apply mathematical operations. This allows
not only the quick inspection and evaluation of multiple images, spectra and 
cubes, but also the production of beautiful multi-color images and animations. 

To improve our understanding of galaxy formation and evolution, we need to 
include models and theoretical knowledge together with observations of galaxy
disks and halos. By fitting and modelling
the observed gas distribution and kinematics of extended galaxy disks, we 
derive their 3-dimensional shapes and rotational velocities (e.g., J\'ozsa 
2007; Kamphuis et al. 2011). Visualisation can then be employed to combine
the actual data with our derived knowledge to re-construct the most likely 
3D representation of each galaxy. By adding time as the fourth dimension one 
can also visualize the evolution of galaxies and the Universe (e.g., see the 
{\em 4D~Universe} visualisation by Dolag et al. 2008). 

A large range of astronomical visualisation packages is available, most of 
which are not discussed here. VO-compatible applications such as {\sc Aladin}, 
{\sc SkyView}, {\sc TopCat}, and {\sc VOPlot} are well suited to viewing
images, spectra and catalogs. Here I briefly highlight a few other 
visualisation tools: 

\begin{figure*}[ht] 
\begin{tabular}{ccc}
 \includegraphics[scale=0.53,angle=0]{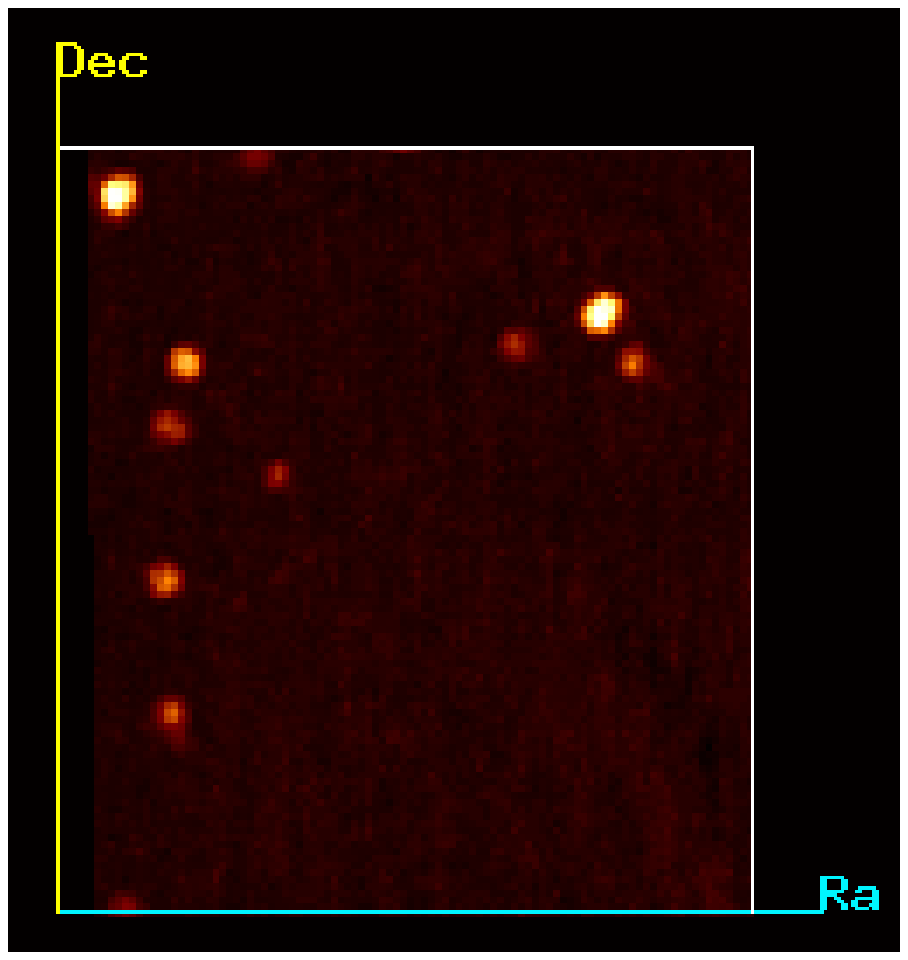}   &
 \includegraphics[scale=0.48,angle=0]{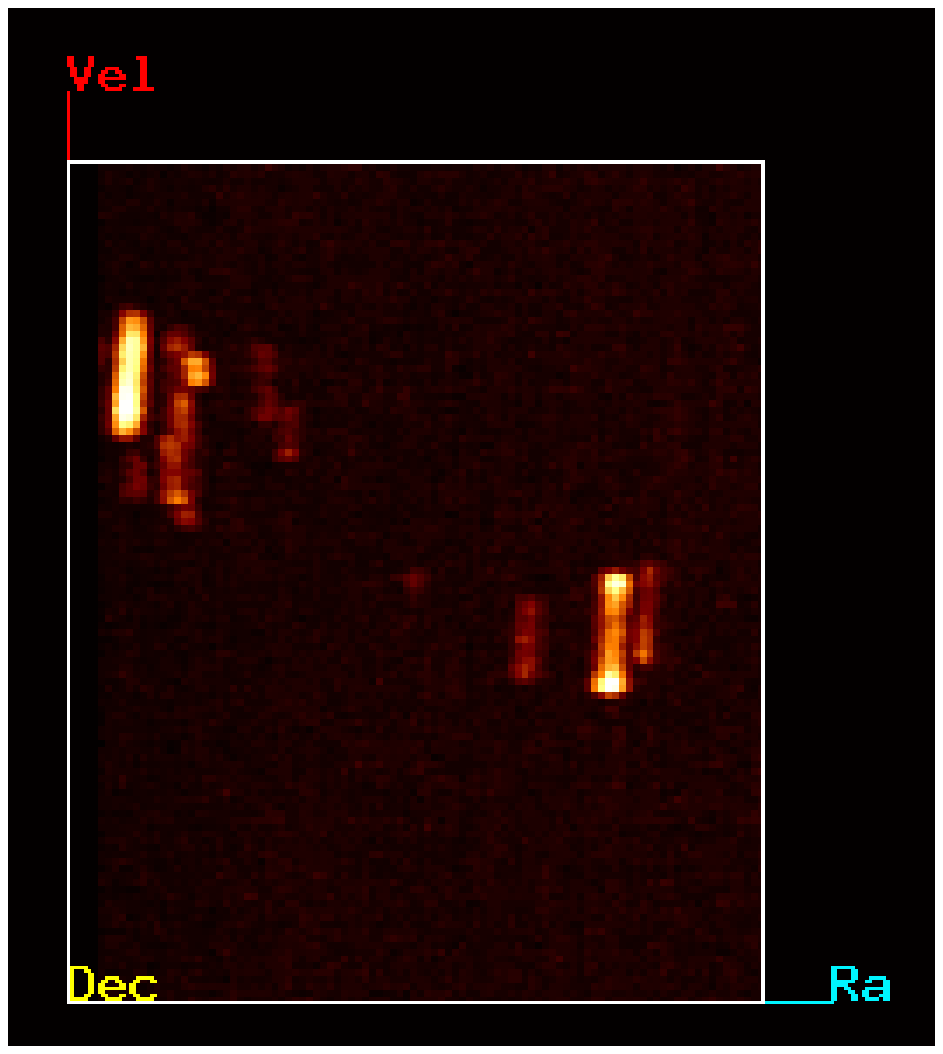}   &
 \includegraphics[scale=0.41,angle=0]{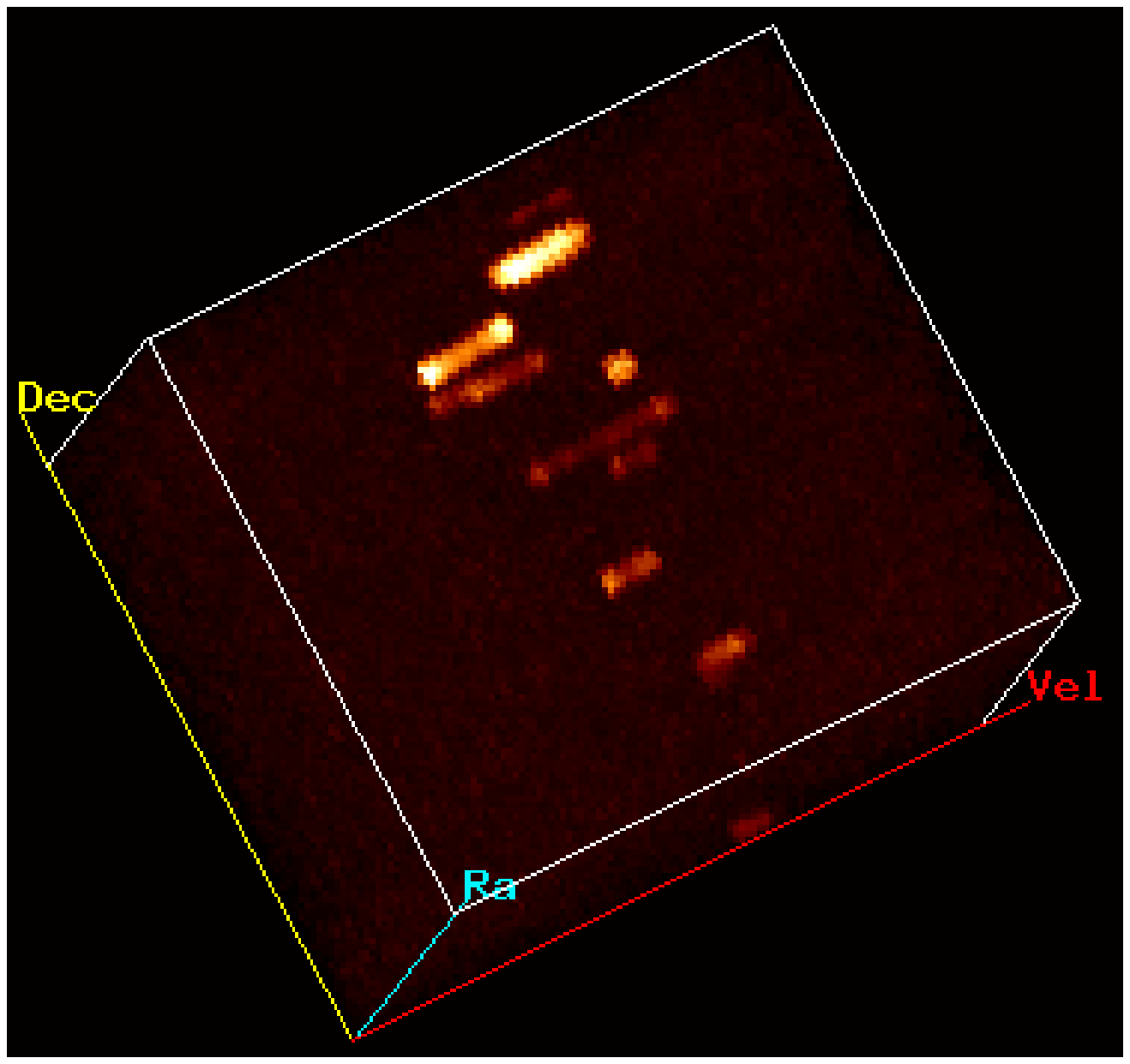}  \\
\end{tabular}
\caption{Visualisation of a HIPASS data cube with the {\sc xray} tool
   in the {\sc KARMA} software package.}
\label{fig:cubes2}
\end{figure*}

\paragraph{\sc KARMA} 
is a widely used, interactive software toolkit\footnote{\em 
  www.atnf.csiro.au/computing/software/karma} for the exploration and analysis
of multi-frequency astrophysical images, spectra and data cubes (Gooch et al. 
1995, 1996). {\sc KARMA} is freely available and highly versatile, including
a diverse range of software tools (e.g., {\sc kvis, xray, koords, kpvslice, 
krenzo, kshell}). The most popular and probably best known tool is {\sc kvis},
capable to visualise multi-wavelength images (FITS format and others) as well
as spectral line data cubes. The {\sc xray} program allows to volume render, 
animate, and explore spectral line (e.g., \HI) cubes (examples are given in 
Fig.~6). The strength of {\sc KARMA} lies in the rapid and intuitive inspection
of small data cubes and images through interactive visualisation. No further 
developments are planned, limiting the usefulness of the package for future 
data sets. The tools (and people) exist to write a much enhanced software 
package, which would be widely used and benefit many researchers.

\paragraph{\sc SPLOTCH}
is a powerful and very flexible ray-tracer software tool\footnote{\em 
   www.mpa-garching.mpg.de/$\sim$kdolag/splotch}
which supports the visualization of large-scale cosmological simulation data 
(Dolag et al. 2008, 2011; Jin et al. 2010). It is publicly available and 
continues to be enhanced. A small team is currently working on supporting the 
visualization of multi-frequency observational data to achieve realistic 3D 
views and fly-throughs of nearby galaxies and galaxy groups. Large-scale 
astrophysical data sets coming from particle-based simulations have been
successfully explored. 

\paragraph{\sc ParaView} is a fully parallel, open-source visualization 
toolkit\footnote{\em www.paraview.org}, used for analyzing and visualizing 
cosmological simulations (see, e.g., Woodring et al. 2011). 

\paragraph{\sc S2Plot} is an advanced 3D plotting library\footnote{\em 
   astronomy.swin.edu.au/s2plot} with support for standard and enhanced 
display devices (Barnes et al. 2006). It provides techniques for displaying 
and interactively exploring astrophysical 3D data sets; for examples see 
Fluke et al. (2010a).

\paragraph{\sc Chromoscope} is an interactive tool\footnote{\em 
   www.chromoscope.net} that facilitates the exploration and comparison of 
multi-wavelength fits images of astronomical data sets on various scales. 
For example applications see Walsh et al. (2012).

\paragraph{\sc VisIVO} is an integrated suite of software tools for the 
visualization of astrophysical data tables (Becciani et al. 2010). Its web 
portal, VisIVOWeb, allows users to create customized views of 3D renderings 
(Costa et al. 2011). In contrast to {\sc SPLOTCH} neither 
{\sc VisIVO}\footnote{\em visivoweb.oact.inaf.it/visivoweb} nor 
{\sc TIPSY}\footnote{\em www-hpcc.astro.washington.edu/tools/tipsy/tipsy.html} 
are designed to lead to ray-tracing like images (Dolag et al. 2008). \\

Specific visualisation challenges for WALLABY are addressed by Fluke et al. 
(2010b), and Hassan \& Fluke (2011). In this {\em PASA Special Issue}
Hassan, Fluke \& Barnes (2012) look into real-time 3D volume rendering of 
large (TBytes) astronomical data cubes. 

\section{Summary \& Outlook} 

The large data volumes (images, cubes, and time series) expected from ASKAP 
and other SKA Pathfinders will require sophisticated source finding algorithms
and visualisation tools. I presented an overview on the current state of 
astronomical source finding with emphasis on spectral line source finding for
extragalactic \HI\ surveys. This snapshot highlights the many developments
underway (e.g., Jurek 2012, Fl\"oer \& Winkel 2012, Allison et al. 2012),
exploring new techniques and testing these comprehensively (Westmeier et al.
2012, Popping et al. 2012). Not only are future source finding algorithms 
required to be highly reliable and complete, they also have to be fast. The
challenge to find faint and/or unusual sources remains. To illustrate the 
difficulty of finding faint \HI\ emission and/or absorption signals in large, 
noise-dominated data cubes, I searched several HIPASS cubes and discussed the 
new detections. The data cubes produced by WALLABY ($\delta < +30\degr$; $z 
< 0.26$), one of several large 21-cm surveys planned for ASKAP, will be so
large that automated source-finding is essential. Visualisation will play a
major role throughout the process of data calibration, source finding and
source analysis. The algorithms used for the visualisation of large data cubes 
(e.g., to enable data quality control and error recognition, to find extended 
source structures, large-scale filaments and voids) are also required to be 
fast, as well as intuitive, interactive, and reliable. I gave an overview of 
some visualisation tools, many of which are also under development to allow 
both the analysis and interpretation of large data volumes, including ASKAP 
21-cm surveys.

\section*{Appendix} 

The new HIPASS detections are briefly described here. Table~2 lists their
\HI\ emission and absorption properties, as fitted with the {\sc MIRIAD}
task {\sc mbspect}. 

\begin{table*} 
\begin{tabular}{llccccl}
\hline
\hline
      & \\
 Name & HIPASS name & \vsys   &  \FHI    & $w_{20}$ & $w_{50}$ & comments \\
      &             & [\kkms] & [Jy\kms] & [\kkms]  & [\kkms]  & \\
      & \\
\hline
      & \\
IC\,2006 + & HIPASS J0354--36a  & 1399 & 9.8 & 254 & 211 & \HI\ emission \\
~~ESO359-G005&                  &      &     &     &     & (group) \\ 
M\,104 (Sombrero) 
           & HIPASS J1239--11   & 1093 & $\sim$16  & 772 & 742 & \HI\ emission \\
           &                    & 1138 & --- &  33 &     & \HI\ abs. \\
      & \\
\hline
      & \\
NGC~5793   & HIPASS J1459--16A  & 3500 & --- & 153 &  88 & \HI\ abs. (1) \\
           &                    & 3540 & --- & 680 &     & \HI\ abs. (1) \\
6dF J1459410.. & HIPASS J1459--16B & 2861 & 5.3 &  90 &  45 & \HI\ emission \\
NGC~5815   & HIPASS J1500--16   & 2999 & 11.3 & 279 & 178 & \HI\ emission \\
      & \\
\hline
      & \\
IC\,5063   & HIPASS J2052--57   & 3421 &  7.5    & 548 & 518 & \HI\ emission \\
           &                    & 2865 &  ---    & 353 & 238 & \HI\ abs. (2) \\
NGC~3801   & HIPASS J1140+17    & 3497 & $\sim$9 & 763 & 146 & \HI\ emission \\
           &                    & 3618 &  ---    & 135 &  76 & \HI\ abs. \\
NGC~612    & HIPASS J0133--36   & 8795 & $\sim$13& 932 & 525 & \HI\ emission \\
           &                    & 8789 &  ---    & 122 &  85 & \HI\ abs. \\
TXS~2226--184 & HIPASS J2229--18& 7302 & ---     & 686 & 612 & \HI\ abs. (3) \\
      & \\
\hline
\hline
      & \\
NGC~4192     & HIPASS J1213+14 & (--132) & (76.5) & (466) && \HI\ emission (4)\\
IC~3105      & HIPASS J1217+12 & --171 & 6.9  & 114 &  88 & \HI\ emission\\
NGC~4294/9   & HIPASS J1221+11 &  +322 & 44.2 & 295 & 261 & \HI\ emission\\
NGC~4523     & HIPASS J1233+15 &  +258 & 20.5 & 133 & 116 & \HI\ emission\\
NGC~4633/4   & HIPASS J1242+14 &  +278 & 11.9 & 240 & 210 & \HI\ emission\\
UGC~08091    & HIPASS J1258+14 &  +213 & 9.0  &  49 &  30 & \HI\ emission\\
      & \\
\hline
\hline
\end{tabular}
\caption{Summary of \HI\ emission and absorption properties for the galaxies 
   discussed in this paper as derived from the respective HIPASS spectra using
   the {\sc MIRIAD} task {\sc mbspect}. For further details see Section~4.1
   and the notes on individual galaxies in the Appendix. ---
   {\bf Notes:} (1) I was able to fit the narrow \HI\ absorption feature in 
   NGC~5793 and give a width estimate of the wide absorption feature. (2) In 
   IC\,5063 I fit the \HI\ absorption between 2700 and 3100\kms. (3) The 
   tentative \HI\ absorption feature in TXS~2226--184 requires confirmation. 
   (4) The \HI\ properties listed for NGC~4192 are from VLA observations by 
   Cayatte et al. (1990).}
\end{table*}

\begin{itemize}
\item {\bf HIPASS J0354--36a}: extended \HI\ emission (see Fig.~1, left) 
  is detected from the region encompassing the galaxies IC\,2006 (\vopt\ 
  = 1381\kms) and ESO359-G005 (\vopt\ = 1399\kms). Both are members of 
  the Fornax cluster which was studied in detail by Waugh et al. (2002). 
  IC\,2006 is an S0 galaxy with a 5-arcmin diameter \HI\ ring (Schweitzer 
  et al. 1989); the dwarf irregular galaxy ESO359-G005 is a gas-rich 
  companion. I measure a total \HI\ flux density of $\sim$9.8 Jy\kms\ 
  (corresponding to an \HI\ mass of $1.6 \times 10^9$\Msun\ assuming a 
  distance of 26 Mpc), about twice the amount detected by Schweitzer et 
  al. (1989). Other prominent S0 galaxies with (partial) \HI\ rings are 
  NGC~1490 (HIPASS J0352--66) and NGC~1533 (HIPASS J0409--56), see
  Oosterloo et al. (2007) and Ryan-Weber et al. (2004), respectively.
\item {\bf HIPASS J1239--11}: wide \HI\ emission (see Fig.~1, right) from 
  the Sombrero galaxy (M\,104, NGC~4594; \vopt\ = 1082\kms) is detected 
  in HIPASS, even though the \HI\ spectrum is strongly affected by baseline 
  ripple. Most prominent are the
  two \HI\ peaks of the double-horn spectrum, separated by nearly 800\kms.
  The \HI\ emission appears to extend along the edge-on dust ring of this 
  beautiful early-type galaxy (Bajaja et al. 1984). There is also a hint of 
  narrow \HI\ absorption ($\sim$1138\kms) against its Sy\,2 nucleus. 
\end{itemize}

The following HIPASS galaxies (see Fig.~2) show both faint \HI\ emission and 
absorption features. Their parametrisation is difficult and remains tentative. 
In most cases, interferometric \HI\ data for these sources exist in the 
literature. For further studies of galaxies with associated \HI\ absorption, 
see van Gorkom et al. (1989), Pihlstr\"om (2001), Taylor et al. (2002), and 
Morganti et al. (2005). 

\begin{itemize}
\item {\bf HIPASS J1459--16A}: wide (680\kms) and narrow (150\kms) \HI\ 
  absorption features are detected towards the well-known water megamaser 
  galaxy NGC~5793 (PKS~B1456--164; \vopt\ = 3491\kms). Figs.~4 \& 5 show the 
  respective HIPASS spectra. The \HI\ velocity range seen in absorption, 
  $\sim$3200 to 3880\kms, is broader than that of the known H$_2$0 masers 
  (Hagiwara et al. 1997) and may indicate a much higher rotational velocity 
  of the nuclear ring and consequently a much larger central mass than 
  previously estimated. This is a spectacular discovery which requires further 
  investigation. There are clear similarities to the \HI\ absorption systems 
  found in PKS~1814--637 and discussed by Morganti et al. (2011). The much 
  narrower but extremely deep \HI\ absorption line seen towards NGC~5793 is 
  discussed by Pihlstr\"om et al. (2000); see also Gardner \& Whiteoak (1986)
  and Koribalski et al. (2004). --- NGC~5793 has several neighbours, two of 
  which are detected in HIPASS: the Sb galaxy NGC~5815 (\vopt\ = 2995\kms) and 
  the dIrr galaxy 6dF J1459410--164235 (\vopt\ = 2857\kms). The E0 galaxy 
  NGC~5796 is probably \HI\ poor.
\item {\bf HIPASS J2052--57}: \HI\ emission and blue-shifted \HI\ absorption 
  is detected towards the Sy\,2 galaxy IC\,5063 (PKS~B2048--572; \vopt\ =
  3402\kms). The latter is indicative of fast gas outflow; for a detailed 
  discussion see Morganti et al. (1998) and Oosterloo et al. (2000). Note 
  that RFI at 1408 MHz ($\sim$2650\kms) also affects the displayed HIPASS 
  spectrum.
\item {\bf HIPASS J1140+17}: wide \HI\ emission and narrow \HI\ absorption is 
  detected towards the FR-I galaxy NGC~3801 (PKS B1137+180; \vopt\ = 3494\kms).
  For a detailed discussion of NGC~3801 and its gas-rich environment see 
  Emonts et al. (2012). Six galaxies and two \HI\ clouds contribute to the 
  emission of HIPASS J1140+17 (from $\sim$3100 to 4000\kms).
\item {\bf HIPASS J0133--36}: wide \HI\ emission and narrow \HI\ absorption 
  is also detected towards the FR-II radio galaxy NGC~612 (PKS~0131--36; 
  \vopt\ = 8925\kms). The mid-point of the \HI\ emission agrees with that of
  the narrow \HI\ absorption ($\sim$8789\kms). For a detailed study see Emonts 
  et al. (2008). We note a $\sim$100\kms\ offset to their \HI\ absorption 
  line measurement. The HIPASS \FHI\ measurement, comprising NGC~612 and 
  neighbouring galaxies, is $\sim$13 Jy\kms, corresponding to $4.7 \times 
  10^{10}$\Msun\ (assuming $D$ = 125 Mpc).
\item {\bf HIPASS J2229--18}: weak \HI\ absorption over a wide velocity range, 
  from $\sim$6900 to 7650\kms, appears to be detected towards the gigamaser 
  galaxy TXS~2226--184 (\vopt\ = 7551\kms). With the majority of the absorption 
  clearly blue-shifted with respect to the systemic velocity, gas outflow is 
  a likely explanation. VLA \HI\ measurements reveal a much narrower (420\kms) 
  absorption feature (Taylor et al. 2002, 2004). Further \HI\ data are needed 
  to confirm this result.
\end{itemize}

HIPASS detections of galaxies in and near the Virgo cluster (see Fig.~3)
are discussed below. Extensive \HI\ studies were carried out by Warmels 
(1988a,b), Cayatte et al. (1990) and Chung et al. (2009).

\begin{itemize}
\item {\bf HIPASS J1213+14}: Cayatte et al. (1990) report on VLA \HI\ 
   observations of the large spiral galaxy M\,98 (NGC~4192). They find the gas 
   distribution to be warped, extending nearly 15 arcmin in diameter over a 
   velocity range from about --360 to +120\kms\ (see also Chung et al. 2009). 
   Assuming a distance of 16~Mpc, we derive an \HI\ mass of $4.6 \times 
   10^9$\Msun. Our HIPASS spectrum is highly contaminated by bright Galactic 
   \HI\ emission in the velocity range from about --100 to +60\kms\ as well 
   as HVCs at higher velocities. 
\item {\bf HIPASS J1217+12}: IC\,3105 is a small edge-on galaxy at a distance
   of 14~Mpc. Our \FHI\ value agrees well with the Arecibo \HI\ measurements by
   Schneider et al. (1990). We derive an \HI\ mass of $3.2 \times 10^8$\Msun. 
\item {\bf HIPASS J1221+11}: NGC~4294/9 is an interacting galaxy pair with
  extended \HI\ emission from both galaxies (Chung et al. 2009). The projected 
  separation between NGC~4294 (\vsys\ = 363\kms) and NGC~4299 (\vsys\ = 
  227\kms) is 5.6 arcmin. Our HIPASS \FHI\ measurement of approximately 
  44 Jy\kms\ for the system (HIPASS J1221+11; Wong et al. 2006) agrees well 
  with the \FHI\ values (27 and 18 Jy\kms) given by Chung et al. from VLA 
  \HI\ data, but is likely an underestimate.
\item {\bf HIPASS J1233+15}: NGC~4523 is Magellanic dwarf galaxy at a distance
   of 17~Mpc. Our \FHI\ value agrees well with previous \HI\ measurements. We 
   derive an \HI\ mass of $1.4 \times 10^9$\Msun. 
\item {\bf HIPASS J1242+14}: NGC~4633/4 is an interacting galaxy pair detected
  with the WSRT in \HI\ emission over a velocity range from --50 to +410\kms\
  by Oosterloo \& Shostak (1993). The projected separation between NGC~4634 
  (\vsys\ = 115\kms) and NGC~4633 (\vsys\ = 297\kms) is 3.8 arcmin. Our HIPASS 
  \FHI\ measurement of 12 Jy\kms\ is slightly lower than the sum of the \FHI\ 
  values (7.4 and 5.8 Jy\kms) given by Oosterloo \& Shostak. Our underestimate 
  is due to Galactic \HI\ emission at velocities below +140\kms\ affecting
  our ability to estimate the blue-shifted \HI\ emission from NGC~4634.
\item {\bf HIPASS J1258+14}: UGC~08091 is a dwarf irregular galaxy at a 
   distance of about 2~Mpc. It lies in the foreground of the Virgo cluster.
   Our \FHI\ value agrees well with previous \HI\ measurements. We 
   derive an \HI\ mass of $8.5 \times 10^6$\Msun. 
\end{itemize}

\section*{Acknowledgements}
\begin{itemize}
\item This research has made extensive use of the NASA /IPAC Extragalactic
      Database (NED) which is operated by the Jet Propulsion Laboratory,
      Caltech, under contract with the National Aeronautics and Space
      Administration.
\item The Digitised Sky Survey was produced by the Space Telescope Science
      Institute (STScI) and is based on photographic data from the UK Schmidt
      Telescope, the Royal Observatory Edinburgh, the UK Science and
      Engineering Research Council, and the Anglo-Australian Observatory.
\item I thank all colleagues for their contributions to this {\em PASA Special
      Issue}. It took much longer than expected to complete all papers, but
      hopefully this issue will provide a comprehensive resource in the field 
      of source finding and visualisation as well as encourage many new ideas
      and further developments.
\item I like to particularly thank the referee for carefully reading the my 
      submitted manuscript and making numerous suggestions which led to a much 
      improved paper. Furthermore, I thank Chris Fluke, Tobias Westmeier, 
      Paolo Serra, Ivy Wong, and Russell Jurek for comments on the content of 
      the paper.
\item I also like to thank Mark Wieringa and the ATNF computing team for 
      many useful updates to the {\sc MIRIAD} software.
\end{itemize}

\end{document}